\documentclass[12pt,preprint]{aastex}

\def\kms{\ifmmode {\rm km\,s}^{-1} \else km\,s$^{-1}$\fi}

\def\simlt{\lower.5ex\hbox{$\; \buildrel < \over \sim \;$}}

\begin{document}

\title{The Twice-Overlooked, Second FR II Broad Absorption Line Quasar LBQS 1138-0126\footnote{Based on observations carried out at the European Southern 
Observatory, Paranal, Chile; program 69.B-0078.}}
\author{Michael S. Brotherton\altaffilmark{1,2}, Scott M. Croom\altaffilmark{3}, 
Carlos De Breuck\altaffilmark{4}, Robert H. Becker\altaffilmark{5,6}, \\ Michael D. Gregg\altaffilmark{5,6}}

\altaffiltext{1}{Kitt Peak National Observatory, National Optical Astronomy Observatories,
950 North Cherry Avenue, P. O. Box 26732, Tucson, AZ 85726; mbrother@noao.edu.
The National Optical Astronomy Observatories are operated by the Association of Universities for Research in Astronomy, Inc., under cooperative agreement with the National Science Foundation.}
\altaffiltext{2}{Department of Physics and Astronomy, University of Wyoming, Laramie, WY 82071}
\altaffiltext{3}{Anglo-Australian Observatory, PO Box 296, Epping, NSW 2121, Australia}
\altaffiltext{4}{Institut d'Astrophysique de Paris, 98bis Boulevard Arago, F-75014 Paris, France}
\altaffiltext{5}{Institute of Geophysics and Planetary Physics, Lawrence Livermore National Laboratory, 7000 East Avenue, P.O. Box 808, L413, Livermore, CA 94550}
\altaffiltext{6}{Physics Department, University of California, Davis, CA 95616}
\begin{abstract}

We report the correct classification of an overlooked Fanaroff-Riley class II 
radio-loud quasar with broad absorption lines, only the second such object
so identified.  The rare properties of this quasar, LBQS 1138$-$0126, are twice overlooked.
First LBQS 1138$-$0126 was found in the Large Bright Quasar Survey but only noted
as a possible broad absorption line quasar without additional follow-up.
Later LBQS 1138$-$0126 was rediscovered and classified as a radio-loud broad 
absorption line quasar but not recognized as an FR II radio source.  
We describe the radio, absorption line, 
and optical polarization properties of LBQS 1138$-$0126 and place it in context with 
respect to related quasars.  In particular, spectropolarimetry shows that 
LBQS~1138$-$0126 has high continuum polarization increasing from 3\% in the red 
(rest-frame 2400 \AA) to over 4\% in the blue (rest-frame 1650 \AA), essentially 
confirming the intrinsic nature of the absorption.  The polarization position angle
rotates from $\sim -30\arcdeg$\ in the red to $\sim 0\arcdeg$\ in the blue;
the radio lobe position angle is $\sim 52\arcdeg$\ for comparison.
LBQS 1138$-$0126 is additionally notable for being one of the most 
radio-loud broad absorption line quasars, and for having low-ionization broad 
absorption lines as well. 

\end{abstract}

\section{Introduction}

Some 10-20\% of optically selected quasars display broad, blueshifted
ultraviolet absorption lines from highly ionized species (e.g., C IV --
HiBAL quasars), with a smaller fraction also showing absorption from very 
low-ionization species (e.g., Mg II -- LoBAL quasars).  
A decade ago, no formally radio-loud broad absorption line 
(BAL) quasars were known despite searching for such objects (e.g., 
Stocke et al. 1992).  This strong anti-correlation seemed to be a clue to 
explaining the apparent bimodal distribution of radio-loudness in quasars 
(which is very much less apparent in new samples selected using deep
radio surveys, see e.g., White et al. 2000).

In the mid 1990s, with the advent of large area, deep radio surveys like
the NVSS (Condon et al. 1998) and FIRST (Becker et al. 1995), radio-loud
BAL quasars began to be found.   Becker et al. (1997) found the first
such source, FIRST J1556+3517, by matching optically red point sources 
with FIRST radio
sources; corrections for dust reddening and the possibility of beaming at
radio frequencies leave this quasar straddling the traditional borderline
between radio-loud and radio-quiet quasars (Najita, Dey, \& Brotherton 2000).
Additional follow-up to the FIRST survey found more radio-loud BAL quasars
by matching radio sources to the APM catalog (Becker et al. 2000; Becker et 
al. 2001) and the Sloan Digital Sky Survey (Menou et al. 2001).  Over 50 BAL 
quasars have now been identified using radio-selection techniques.

The largest and brightest sample of radio-selected BAL quasars is that of
Becker et al. (2000), which comprises $\sim$27 BAL quasars from the FIRST Bright
Quasar Survey (White et al. 2000).  Their radio properties can tell us
some things that cannot be learned from optically selected BAL quasars.
First, a wide range of radio spectral indices are present, including both flat and
steep spectra; unified radio models (e.g., Orr \& Browne 1982) would
indicate that therefore a range of orientations are present.  Second, the radio
sources are almost all compact (90\%), whereas a matched parent population from
the FIRST Bright Quasar Survey (FBQS) consists of only 60\% compact sources.  
This is similarly contrary to the idea that BAL quasars are simply normal quasars 
seen edge-on.  Properties at other wavebands are also inconsistent with
BAL quasars being edge-on quasars (e.g., Brandt \& Gallagher 2000).

Were it not for these results, the discovery of a powerful edge-on
Fanaroff-Riley type II (FR II; Fanaraoff \& Riley 1974) radio source
that is also BAL quasar, FIRST J101614.3+520916 (hereafter FIRST J1016+5209,
Gregg et al. 2000), might have been taken as supporting evidence for 
an edge-on geometry.  The once popular notion of such an edge-on geometry 
was motivated by spectropolarimetry (e.g., Ogle et al. 1999) and by theoretical
expectations for winds arising from accretion disks (e.g., Murray et al. 1995;
Konigl \& Kartje 1994).  Instead, Becker et al. (2000) and Gregg et al. (2000) 
interpret BAL quasar radio properties to indicate that BAL quasars are young, 
evolving quasars, moving from dusty objects with high covering fractions and
high accretion rates to becoming more typical quasars.

Brotherton et al. (1998) found five radio-loud BAL quasars by matching
color-selected quasar candidates from the 2dF Quasar Survey (e.g., 
Croom et al. 2001) against radio sources in the NVSS Survey.  
These were of particular interest because their blue colors ensured 
minimal dust-reddening which can artificially enhance the radio-loudness 
by making an object appear optically weak relative to its radio emission.  
Also of interest was that the observed targets were taken from a large 
optically selected sample, making possible statistical comparison with 
the Large Bright Quasar Survey (LBQS; Foltz et al. 1989; Hewett et al. 
1991; Chaffee et al. 1991; Morris et al. 1991; Francis et al. 1992).
The 2dF-NVSS cross-match indicates that approximately 1 in 500 
bright blue quasars should be seen in an observed-frame optical
spectrum as a radio-loud BAL quasar.  With the LBQS coming in at $\sim$1000 
quasars it would be likely to find 1 or 2 such objects, but not too
surprising to not find any.

An error labeling the coordinates of one of the five radio-loud BAL quasars
from Brotherton et al. (1998) prevented identification with the correct
radio source (Brotherton et al. 2002).  Likewise, this error prevented
identification with a previously discovered quasar LBQS 1138$-$0126
(Hewett et al. 1991), which had only been noted as being a possible BAL 
quasar but had not been followed-up in either the optical or radio.
We show in this paper that LBQS 1138$-$0126 is not only a BAL quasar,
it is also a powerful FR II radio source, only the second such BAL
quasar found.

\section{Data}

In addition to the 4000--9000 \AA\ spectrum presented by
Brotherton et al. (1998), the original LBQS spectrum
(Hewett et al. 1991) is also available which covers shorter wavelengths,
including the C~IV $\lambda$1549 transition.  Additionally,
radio data is publicly available from several surveys,
including radio maps from the FIRST Survey (Becker, White,
\& Helfand 1995).  Finally we describe new spectropolarimetry
obtained from European Southern Observatory's Very Large
Telescope (VLT).

\subsection{Radio Observations}

A search on the NASA Extragalactic Database (NED\footnote{The 
NASA/IPAC Extragalactic Database (NED) is operated by the Jet Propulsion 
Laboratory, California Institute of Technology, under contract with the 
National Aeronautics and Space Administration.}) shows that LBQS~1138-0126 
is present in several radio surveys in addition to the NVSS,
which indicates a radio flux density of 252 mJy at 1.4 GHz and a 
radio-loudness of log R* = 3.0 (Brotherton et al. 2002).  This
already makes LBQS~1138-0126 competitive for the title of the most
radio-loud BAL quasar known.

In the Texas 365 MHz Survey (Douglas et al. 1996), LBQS~1138-0126
is reported to have a flux density of $658 \pm 49$\ mJy.
In the Parkes-MIT-NRAO (PMN) Survey (Griffith et al. 1995), LBQS~1138-0126
is reported to have a flux density of $69 \pm 11$\ mJy at 4.85 GHz.
The flux measurements at the three frequencies indicate a radio
spectral index of $\alpha = -0.7$\ ($S_{nu} \propto \nu^{\alpha}$)
from 365 MHz to 1.4 GHz, steepening to $\alpha = -1$\ from 1.4 GHz to
4.85 GHz.   LBQS~1138-0126 is a strong, steep-spectrum radio source 
consistent with optically thin synchrotron from a lobe-dominated FR II source.

The high resolution of the 1.4 GHz FIRST survey (Becker et al. 1995) 
provides morphological information.  We have used the FIRST Survey cutout 
server\footnote{http://sundog.stsci.edu.}
to extract an image of LBQS~1138-0126 at $\sim$5\arcsec\
resolution.  Fig. 1 shows this image as a contour map.  We clearly
see a double-lobed radio structure at a position angle $\theta \sim 52\arcdeg$,
with a bridge of radio flux between.  A spur from the NE source into this
bridge corresponds to the marked optical position of LBQS~1138-0126,
although from this map we cannot reliably extract a distinct core flux.
The core flux does appear to be no more than a tenth of the lobe flux,
and probably significantly less, making the source ``edge-on'' in unified
radio schemes (e.g., Orr \& Browne 1982).

\subsection{Spectropolarimetry}

We obtained spectropolarimetric observations of LBQS~1138-0126 on UT
2002 May 8 using the PMOS mode of the FORS1 spectrograph (Appenzeller
et al 1992) on the Melipal unit of the VLT. Conditions were nearly
photometric with $\sim 0.7\arcsec$ seeing. The observations were split
into four 300s exposures, each at a different orientation of the
half-wave plate (0\arcdeg, 22.5\arcdeg, 45\arcdeg, 67.5\arcdeg).  
We used the 600B grism with a 1.0$\arcsec$\ wide slit oriented North-South, 
resulting in a spectral resolution of 5.0~\AA~(FWHM).

We reduced the data using the standard IRAF procedures. We extracted
the spectra with an identical $3\farcs6$\ wide aperture for the o and
e-rays, and re-sampled all 8 individual spectra (o and e-rays for the
4 half-wave plate positions) to the same linear dispersion
(1.166~\AA/pix) in order to calculate the polarization in identical
spectral bins. We used the procedures of Vernet (2001; see also 
Vernet \& Cimatti 2001), which are
based on the method described by Cohen et al. (1997).  We checked the
polarization angle offset between the half-wave plate coordinate and
the sky coordinates against values obtained for the polarized standard
stars Vela1 and Hiltner 652; our values are within $<1\arcdeg$ from the
published values, and the polarization percentage within 0.1\%.
Figure 2 shows our results.

We explicitly note that the broadband optical polarization results 
for this quasar (called UN J1141-0141 as in Brotherton et al. 1998) 
reported by Lamy \& Hutsem{\' e}kers (2000) are incorrect
since they observed the wrong quasar because of the originally
incorrect coordinates (Brotherton et al. 1998; Brotherton et al. 2002).

\section{Analysis \& Results}

\subsection{BALs in the LBQS 1138$-$0126 Spectrum}

Brotherton et al. (1998) classified LBQS 1138$-$0126 as a BAL quasar
on the basis of broad but shallow absorption troughs from Al III and
Mg II.  Typically these low-ionization BALs, when present, are 
weaker and lower velocity than high-ionization BALs (e.g., Voit et al. 1993).
The original LBQS spectrum and our new VLT spectrum cover the C IV
$\lambda$1549 region (not present with the Brotherton et al. (1998) 
wavelength coverage) and reveal a significantly deeper BAL trough (Fig. 3).

The original LBQS spectrum has higher signal-to-noise ratio (SNR) than the
VLT spectrum in the immediate region of the C IV $\lambda$1549 BAL,
although the VLT spectrum has the best SNR at 
longer wavelengths that includes the Al III $\lambda$1863 BAL.
The C IV $\lambda$1549 BAL is present in both HiBAL quasars and LoBAL 
quasars, and provides the most commonly measured standard for comparison
of absorption line properties.  Weymann et al. (1991) defined the ``BALnicity 
Index'' (BI) -- a complex measurement of the detachment, depth, and velocity
span of an absorption feature -- in order to distinguish real BALs from 
blends of associated absorbers (especially) in low-resolution, low-SNR spectra.
In recent years detailed studies of individual objects have shown 
that a significant number of absorbed quasars with zero BALnicity nevertheless 
possess intrinsic outflows (e.g., Becker et al. 2000).  
For this reason Hall et al. (2002) have created a new index similar to
the BI, called the ``Absorption Index'' or AI,  that provides a wider measurement 
range to include such lower velocity outflows. 

We find that the C IV $\lambda$1549 BAL trough, based on the LBQS spectrum,
possesses a small but positive Balnicity Index, BI = $900 \pm 300$ \kms. 
The uncertainty is quite large given the low SNR, and depends critically on 
continuum choice and assumed redshift (we assume $z=1.266$).  The uncertainty
we quote indicates the range of values we find for different assumptions
rather than formal fitting errors.  The Absorption Index AI = $3000 \pm 300$ \kms.
The conservative definition of BI would indicate that we certainly have
a BAL quasar in LBQS 1138$-$0126, and this is supported by the deep C IV 
trough and velocity span of some $\sim 5000$ \kms detached $\sim 2000$ \kms
from the C IV $\lambda$1549 peak.  The low-ionization BAL troughs (Al III,
Mg II) cover a very similar velocity range (approximately $-$2000 to $-$8000 \kms) 
but are much shallower.

If there is any doubt regarding the BAL quasar designation for LBQS 1138$-$0126,
they should be dispelled by polarization results.  Only about 1 in a
hundred optically selected quasars without intrinsic absorption shows any 
significant optical polarization greater than 1\% (e.g., Berriman et al. 1990).  
The frequency is similar for radio-loud quasars with extended, lobe-dominated
steep-spectrum radio structures (e.g., Visvanathan \& Wills 1998 and 
references within).  BAL quasars, especially those showing absorption from
low-ionization species and red colors, are significantly more often highly 
polarized than unabsorbed quasars (Brotherton et al. 2001a and references therein).  
The only other quasar class exhibiting such high
levels of continuum polarization are beamed radio-loud quasars (blazars) 
which possess a large contribution of optical synchrotron emission; the radio
structure of LBQS 1138$-$0126 rules out a blazar designation.

\subsection{Reddening in LBQS 1138$-$0126 and Radio-Loudness}

Despite the fact that LBQS 1138$-$0126 has an ultraviolet excess
($U - B_j = -0.7$), the continuum shape is clearly  
redder than the average radio-selected quasar.  Its selection
was assisted by the presence of C IV $\lambda$1549 emission 
in the U-band.  The Galactic reddening is only E(B-V) = 0.017 mag in 
this direction (Schlegel et al. 1999), so local Galactic dust is not the cause.  
For the LBQS, Sprayberry \& Foltz (1992) showed that LoBAL quasars
were on average reddened by $E(B-V)$ = 0.1 mag for a Small Magellanic Cloud extinction 
law (Prevot et al. 1984) compared to the LBQS as a whole.
Comparison to the radio-selected composite FBQS spectrum (Brotherton et al. 2001b) 
indicates that the continuum of LBQS 1138$-$0126 is consistent with intrinsic 
reddening (at the redshift of the quasar) of 0.35 mag A$_V$, for a Small Magellanic 
Cloud extinction law which well matches the average result of Sprayberry \& Foltz (1992).
The effective observed B-band flux would then be enhanced
by a factor of $\sim$2.8 to correct for this level of extinction.
The intrinsic radio-loudness would then decrease,
because of the intrinsically brighter optical continuum, to
log R* = 2.5.  This value would still place LBQS 1138$-$0126
among the most radio-loud BAL quasars known.

\section{Discussion}

It is clear that LBQS 1138$-$0126 is a powerful
double-lobed FR~II radio source, and that the optical/UV
spectrum contains both high-ionization and low-ionization
BALs.  The high continuum polarization makes a very strong
statistical argument that the absorption lines are broad
and intrinsic and we are not being misled by the 
low-resolution of our spectra.  These properties
lead to the conclusion that LBQS 1138$-$0126 is an FR II
BAL quasar, only the second such object so identified after 
FIRST J1016+5209 (Gregg et al. 2000).

We also note the existence of the less luminous, low-redshift quasar
PKS 1004+13 (Wills et al. 1999) that is a strong candidate
for having BALs and is soon to have a definitive UV spectrum
obtained with the Hubble Space Telescope.
One other BAL quasar from Brotherton et al. (1998),
UN J1053$-$0058, is seen to be a very core-dominated radio triple
in FIRST Survey images.  The spatial exent of the triple
is approximately 45\arcsec, corresponding to a projected 
size of 550 kpc at $z=1.55$ for our adopted cosmology.
Hutsem{\'e}kers \& Lamy, H. (2000) report an optical 
polarization of 1.89\% with a position angle some 20-30\arcdeg
from being perpendicular to the radio axis.  The large
radio core dominance (an observed core-to-extended 1.4GHz
flux ratio of about 12) would indicate that the radio
core, and the radio-loudness, is artificially enhanced by 
relativistic beaming.  Therefore UN J1053$-$0058 is 
perhaps best identified as a beamed radio-quiet quasar
rather than as an intrinsically luminous radio-loud 
quasar (see Falcke, Patnaik, \& Sherwood 1996; 
Falcke, Sherwood, \& Patnaik 1996).  We focus the 
remaining discussion on quasars like LBQS 1138$-$0126 and 
FIRST J1016+5209.

How similar are LBQS 1138$-$0126 and FIRST J1016+5209?
Table 1 compares their properties.  The similar small BIs, 
modest reddening, high polarization and radio luminosity
in particular stand out.  Also similar is the fact that
both have optical polarization position angles intermediate
between those parallel or perpendicular to the
large-scale radio structure.  The comparison of the angular
size is strongly affected by adopted cosmology although
both sources appear to be relatively large (at least several
hundred kpc).  Significant differences include the presence of 
low-ionization BALs in LBQS 1138$-$0126 but not
FIRST J1016+5209, and the smooth single trough
structure of LBQS 1138$-$0126 compared to the jagged 
multiple trough structure in the BALs of FIRST J1016+5209 which
also spans a much larger velocity range.  We note that 
among radio-quiet BAL quasars there exists a wide range of
trough structures which have so far not been found to 
clearly correlate with other properties.

Gregg et al. (2000) proposed and discussed an evolutionary
sequence for LoBAL quasars to evolve into HiBAL quasars
and then radio-loud quasars with associated absorption that
is the remnants of the once smooth BAL outflow. 
The formally radio-loud BAL quasars from Becker et al. (2000)
with compact radio structures would be the youngest
(most recently fueled) sources, perhaps themselves 
closely related to giga-Hertz peaked sources (GPS) and 
compact steep spectrum (CSS) radio-loud quasars (e.g., O'Dea 1998).
Next in the progression is LBQS 1138$-$0126, in which the radio jet
has escaped the nuclear regions but low-ionization BALs
are still visible and dust is still present along the line of
sight.  FIRST J1016+5209 is the next stage as the LoBAL
features vanish and the HiBAL features begin to break up.
The final stage before being regarded as a normal, unabsorbed
radio-loud quasar is represented by PKS 1157+014, an object once put forward
and rejected as a radio-loud BAL quasar for which variability
has provided proof of the intrinsic nature of its rather
narrow absorption features (Aldcroft, Bechtold, \& Foltz 1998).

The polarization properties of LBQS 1138$-$0126 are similar
to those of other highly polarized BAL quasars (e.g., Ogle
et al. 1999).  The polarization mechanism is widely thought
to arise from asymmetric scattering as other likely mechanisms
can be ruled out with high-quality spectropolarimetry 
for individual objects.  In particular, the rise in the polarization
toward shorter wavelengths is very common and can be 
attributed to a decreasing amount of dilution from direct
light more reddened than the scattered light path or 
as a signature of dust scattering (e.g., Hines et al. 2001).
Less commonly seen in polarized BAL quasars is a significant
rotation of the polarization position angle, although it appear
(e.g., QSO 2359$-$1241, Brotherton et al. 2001a).  Such a rotation 
can be ascribed to multiple
scattering light paths with different amounts of reddening
or different polarization efficiencies as a function of
wavelength (e.g., in the case of dust scattering).  Such
complex scattering geometries might be expected for young,
highly accreting objects with large covering fractions, 
as opposed to a simpler scattering geometry as seen in 
edge-on Seyfert 2 and radio galaxies for which the polarization
position angle is perpendicular to system (jet) axes
(Antonucci 1993). 

LBQS 1138$-$0126 is not especially red or faint and
was easily found using optical selection techniques.
The radio emissions are strong and detected in multiple
surveys.  The key elements in making the classification
of a radio-loud FR II BAL quasar are having an optical
spectrum covering the C IV $\lambda$1549 region and 
matching optical properties with radio properties.  
Given the statistics of Brotherton et al. (1998) and
Menou et al. (2000) selecting BAL quasars by optical
techniques and matching to radio surveys, the SDSS
should turn up an entire population of quasars similar
to LBQS 1138$-$0126.  The small number statistics 
and the possible existence of as yet unidentified 
biases make a precise estimate impossible, but we 
might expect 100-200 FR II BAL quasars to be identifiable
from combining information from SDSS discovery spectra 
and the FIRST Survey.  This might seem like a large number 
of sources, but these still appear to be rare objects requiring 
very large surveys to discover in significant numbers.
Such a sample would permit quantitative tests and 
development of the evolutionary hypothesis of 
Gregg et al. (2000).

\section{Summary}

The properties of LBQS 1138$-$0126, now properly recognized,
mark it as the second FR II BAL quasar so far discovered
after FIRST 1016+5209.  This quasar shows both low and 
high-ionization BALs as well as the high continuum polarization 
(3-4\%) characteristic of LoBAL quasars.  The position
angle of the luminous radio lobes differs from the polarization
position angle by 50-70\arcdeg.  Modest reddening of the 
optical/ultraviolet continuum reduces the radio-loudness
from log R* = 3.0 to log R* = 2.5, still one of the most
radio-loud BAL quasars known and one of only two confirmed
BAL quasars with a large double-lobed radio structure.

\acknowledgments

We thank Paul Francis for providing the original LBQS spectrum, and
Joel Vernet for providing his IDL spectropolarimetric reduction code.
We also thank Bev Wills for pointing out the radio structure of 
UN J1053$-$0058.
A portion of this work has been performed under the auspices of the 
U.S. Department of Energy, National Nuclear Security Administration by the
University of California, Lawrence Livermore National Laboratory under 
Contract W-7405-ENG-48.
This research has made use of the NASA/IPAC Extragalactic Database (NED) 
which is operated by the Jet Propulsion Laboratory, California Institute 
of Technology, under contract with the National Aeronautics and 
Space Administration.  
This work was supported by a Marie Curie Fellowship of the European
Community programme 'Improving Human Research Potential and the
Socio-Economic Knowledge Base' under contract number HPMF-CT-2000-00721.


\begin{deluxetable}{lcc}
\tabletypesize{\footnotesize}
\tablewidth{0pt}
\tablecaption{Properties of FR II BAL Quasars\tablenotemark{a}}
\tablehead{
Parameter & LBQS 1138$-$0126 & FIRST 1016+5209\tablenotemark{b}
}
\startdata
R.A. (J2000) & 11 41 11.56 & 10 16 14.3 \\
Decl. (J2000) & $-$01 43 07.7 & +52 09 16 \\
$z$ & 1.266 & 2.455\\
$B$\tablenotemark{c} & 18.56 & 20.2 \\
$R$\tablenotemark{d} & 17.74 & 18.6 \\
$S_{365 MHz}$\ (mJy) & 658 & 846\tablenotemark{e} \\
$S_{1.4 GHz}$\ (mJy) & 252 & 131.1 \\
$S_{4.8 GHz}$\ (mJy) &  69 & 44 \\
Lobe peak-to-peak distance (arcsec) & 21 & 45 \\
Total projected size (kpc) & 250 & 600 \\ 
Lobe-lobe axis P.A. (deg) & 52 & 146 \\
Continuum Pol. P.A. (deg) & 0 to $-$30& 85 to 75 \\
Continuum Pol. (\%) & 4 to 3 & 2.5 to 2\\
C IV $\lambda$1549 BI (\kms) & 900 & 2400 \\
Intrinsic $E(B-V)$\ (mag) & 0.11 & 0.27 \\
$M_B$ & $-26.4$\ ($-$26.9) & $-25.7$\ ($-26.8$) \\
Log ($L_{1.4 GHz}$) (ergs s$^{-1}$ Hz$^{-1}$) & 34.4 & 34.5 \\
Log R* & 3.0 (2.5) & 3.4 (2.7) \\
\enddata
\label{table1}
\tablenotetext{a}{From Brotherton et al. (2002) or measured here,
for $H_o = 50$\ \kms Mpc$^{-1}$ and $q_o = 0$.  Quantities in 
parentheses correspond to dereddened values.}
\tablenotetext{b}{Taken from Gregg et al. (2000) and converted to
our adopted cosmology.}
\tablenotetext{c}{$B_j$ in the case of LBQS 1138$-$0126, POSS $O$\ in the
case of FIRST 1016+5209.}
\tablenotetext{d}{POSS $E$\ in the case of  FIRST 1016+5209.}
\tablenotetext{e}{Also from the Texas Survey, Douglas et al. (1996).}
\end{deluxetable}

\begin{figure}
\plotone{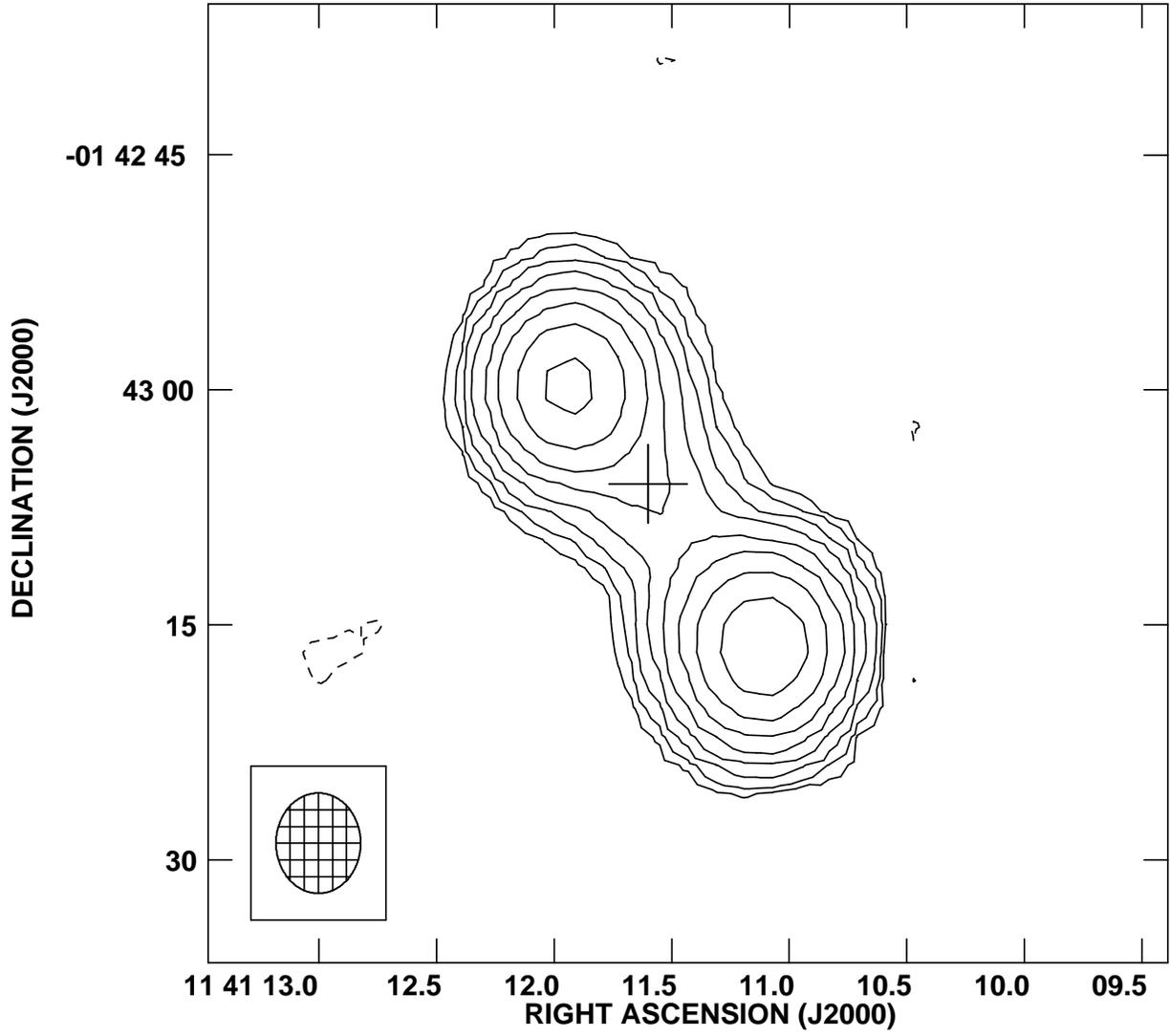}
\caption{FIRST Survey 2 $\times$ 2 arcminute cut-out.
Displayed solid contours start at 0.5 mJy and increase by factors of 2.
The dashed contours are at $-$0.5 mJy.  The beamsize is shown at the 
bottom left.  The cross marks the optical position of the quasar.
}
\end{figure}

\clearpage

\begin{figure}
\plotone{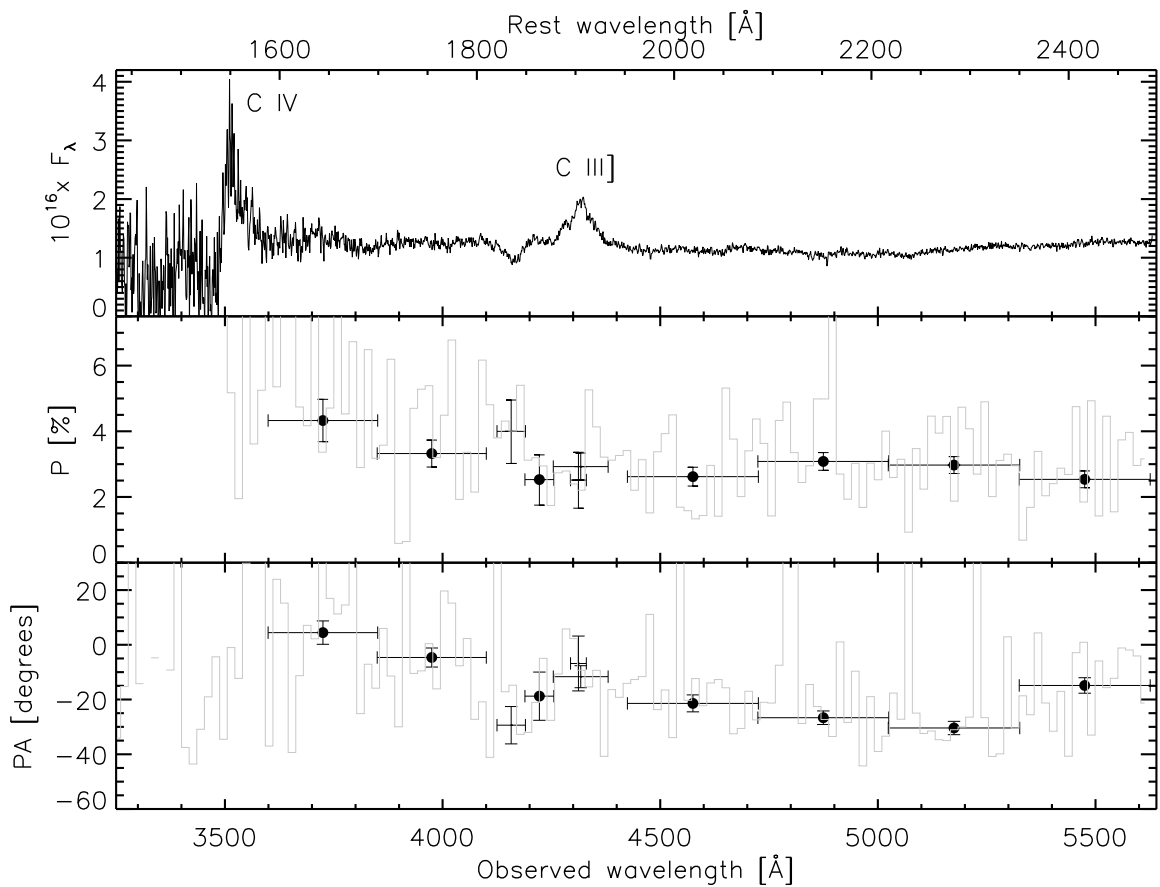}
\caption{
VLT spectropolarimetry of LBQS 1138$-$0126.  The top panel shows the
total flux spectrum and labels the two prominent emission lines.
The middle panel shows the linear polarization in wide bins covering
the continuum and narrower bins across the emission and trough structures.
The bottom panel shows the position angle of the electric vector
using the same bins as the middle panel.  Continuum bins are marked by 
circles, line bins only by error bars -- all error bars are 1 $\sigma$.
The gray histograms in the bottom two panels show the results for a fine
binning.  Both observed-frame and rest-frame (using $z=1.266$) wavelength 
scales are provided.
}
\end{figure}

\begin{figure} 
\plotone{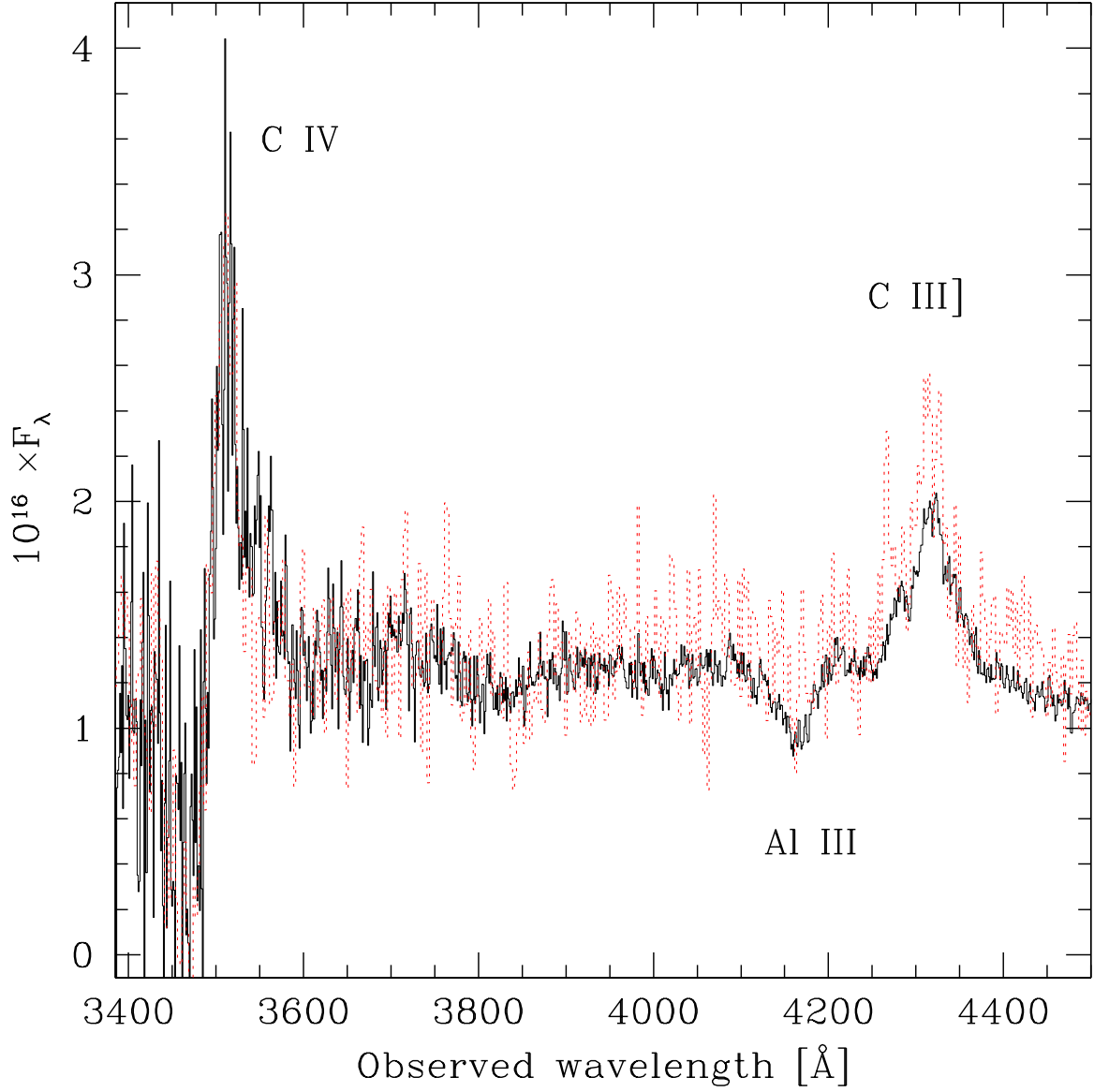} 
\caption{
Comparison of total flux spectroscopy 
(ergs s$^{-1}$ cm$^{-2}$ \AA$^{-1}$ $\times 10^{16}$)
between our new VLT spectrum (solid line) and the original LBQS 
discovery spectrum (dotted/red line).  
}
\end{figure}

\end{document}